# Variable Cutoff Frequency FIR Filters: A Survey

Author list: Sumedh Dhabu[1], Abhishek Ambede[1], Smitha K. G.[1], Sumit Darak[2], A. P. Vinod[3]

[1] Nanyang technological University, Singapore, [2] IIIT-Delhi, India, [3]IIT-Palakkad, India

E-mail: sdhabu@ntu.edu.sg, abhisheka@ntu.edu.sg, smitha@ntu.edu.sg, sumit@iiitd.ac.in, vinod@iitpkd.ac.in

**Abstract**

Many signal processing applications require digital filters with variable frequency characteristics, especially the filters with variable bandwidth. Due to their linear phase and inherent stability, variable bandwidth finite impulse response (FIR) filters are the popular choice in majority of the applications. Once a variable cutoff frequency (VCF) FIR lowpass filter is designed, variable bandwidth bandpass / highpass / bandstop filters and reconfigurable filter banks can be realized from the same. In this paper, we present a comprehensive review of the existing variable cutoff frequency FIR filter design techniques, including the developments in the recent two decades. We provide the basic concepts, design and architectural details for each of these techniques and the major developments / incremental works thereof. Qualitative as well as quantitative comparisons are provided to assist the reader in choosing the most suitable VCF filter design technique for a particular application.

**Keywords**

FIR filters, reconfigurable filters, variable digital filters

## 1. Introduction

Compared to analog filters, digital filters have advantage of higher accuracy, exact linear-phase (if desired), time-invariant performance (analog filter performance changes due to component drift), relatively smaller silicon area etc. These advantages, coupled with the advent of VLSI technology, have resulted in ubiquitous presence of digital filters in almost all the signal processing applications with a few exceptions (such as wideband filtering in radio frequency range, which is not feasible due to the bottleneck at the speed of analog-to-digital converters). Digital filters with variable frequency response characteristics are required for numerous critical applications in the fields of digital communications, audio signal processing, biomedical signal processing etc. [1, 2]. In digital communications domain alone, digital filters with variable frequency response characteristics can be used for spectrum analysis, spectrum shaping, channelization, and receiver synchronization. For e.g., a digital filter having tuneable passband and transition bandwidths is used for channel extraction in a multi-standard wireless communication receiver, as it needs to be interoperable with multiple communication standards having distinct bandwidth specifications. Alternatively, a digital modem uses a digital filter to implement variable fractional delays. In general, a digital filter whose frequency response can be changed on-the-fly is called a variable digital filter. Even though such a definition means that all the parameters related to the frequency response should be variable, in practice, majority of the applications demand only a lowpass filter with variable cutoff frequency ($f_c$), satisfying the desired minimum specifications of transition bandwidth (*tbw*), passband ripple ($\delta_p$) and stopband ripple ($\delta_s$). Therefore, the VCF (variable cutoff frequency) filter design task can be simplified further, by specifying the maximum limit on *tbw*, $\delta_p$, and $\delta_s$, and thus, only one parameter related to the frequency response, i.e., $f_c$, needs to be variable. These parameters related to design of a VCF filter are illustrated in Fig. 1.

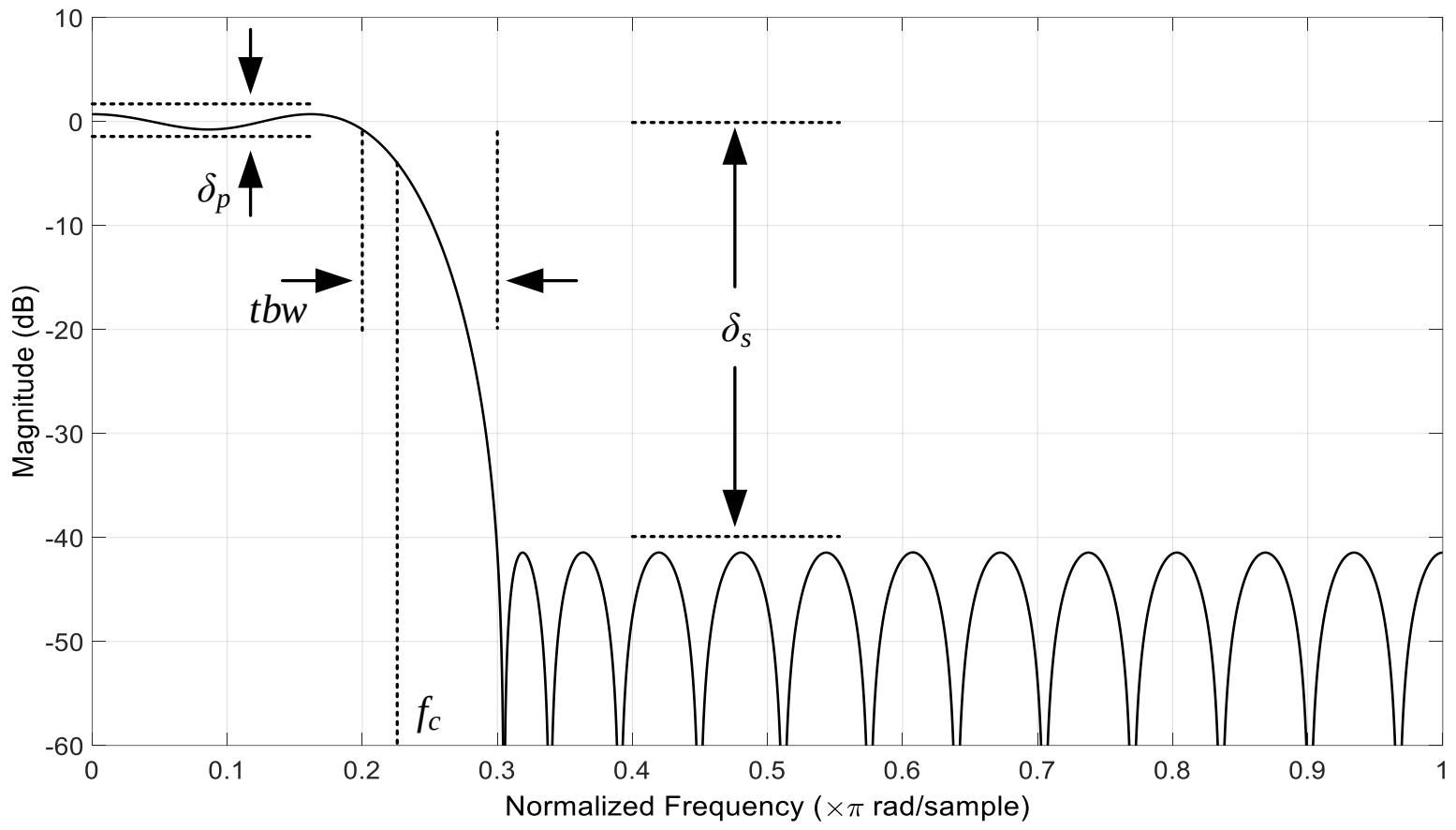

Fig. 1. Magnitude response characteristics of lowpass filter and design parameters.

Depending on the application requirements, a VCF filter can be an infinite impulse response (IIR) filter (a recursive filter) or a finite impulse response (FIR) filter (a non-recursive filter) [3]. For instance, for energy-detection based spectrum sensing in cognitive radio, only the magnitude of a particular frequency component is of interest, and therefore, IIR filter provides a computationally efficient alternative to an FIR filter. Whereas, for the channelization task (extraction of individual frequency band from wideband signal), a linear-phase response is required, and therefore, an FIR filter is used. Besides linear phase, FIR filters are inherently stable as opposed to IIR filters. They provide a constant group delay for all the frequencies and have better performance with fixed-point implementation when compared to IIR filters. These advantages make them a popular choice in many applications. A previous work on the review of VCF FIR and IIR filters can be found in [2]. On the contrary to the conclusion of this review in [2], due to opening up of new research avenues (e.g., cognitive radios, new waveforms for 5G communications), a significant number of new noteworthy publications has been witnessed in the field of design and implementation of one-dimensional variable filters.

Cognitive radio has become a hot topic of research in last two decades. Reconfigurable digital front-end in the cognitive radio requires an area-efficient reconfigurable linear-phase filter in order to perform channelization and spectrum sensing [68]. Consequently, a lot of novel design approaches have been proposed for the design of area-efficient, low complexity VCF FIR filters. In this paper, we present a comprehensive review of all the VCF FIR filter design techniques, including these recent developments. We present their hardware implementation architectures and qualitative comparison in terms of advantages and limitations in a broader sense. This paper is intended to serve as a tool to initiate the reader into the field of VCF one-dimensional FIR filter design techniques, finer details of which can then be referred to from related literature. Many of these techniques can be easily adapted for designing one-dimensional and two-dimensional reconfigurable filter banks and VCF IIR filters, after appropriate modifications (and considerations to maintain the stability in case of IIR filters). Interested readers are advised to refer to individual papers for detailed information.

The frequency response of a digital filter depends on its impulse response, and therefore, by modifying the impulse response of a filter structure, a VCF filter is obtained. To understand various VCF filter design techniques, let us first consider the FIR filter architecture. Compared to the direct-form structure, the transposed direct-form structure is widely used to implement FIR filters in hardware, as it can achieve higher operating frequency due to inherent pipelining. For symmetric coefficient FIR filters, only about half of the total number of coefficients need to be implemented. Fig. 2 shows a transposed direct-form structure for implementing an FIR filter of even order $N$, with coefficients denoted by $h_0$, $h_1$ …$h_{N/2}$. Here, the coefficients are fixed-valued and the second multiplicand (i.e. input sample) is variable. (Fixed-coefficient multipliers are represented by triangles in this paper.) Each block with $z^{-1}$ provides a unit delay. This filter structure outputs a fixed impulse response, corresponding to a fixed frequency response (i.e., one particular cutoff frequency). Building blocks of this filter structure (such as coefficient multiplier, delay etc.) can be made variable, or substituted by some other variable structure, in order to obtain a variable impulse response. Depending on whether the coefficients $h_0$, $h_1$, … etc. are variable or not, the VCF filters can be classified as variable-coefficient VCF filters and fixed-coefficient VCF filters.

This paper presents an exhaustive survey of the VCF FIR filter design techniques. The rest of the paper is organized as follows – Section 2 presents variable-coefficient VCF filters followed by a detailed discussion of fixed-coefficient VCF filters in Section 3. Design concepts, hardware implementation architectures and qualitative as well as illustrative quantitative comparisons are provided for the different VCF filter design techniques. Detailed theory and mathematics are deliberately omitted as the purpose of this review is to be an introductory article and the interested readers are encouraged to look up the individual references for the same.

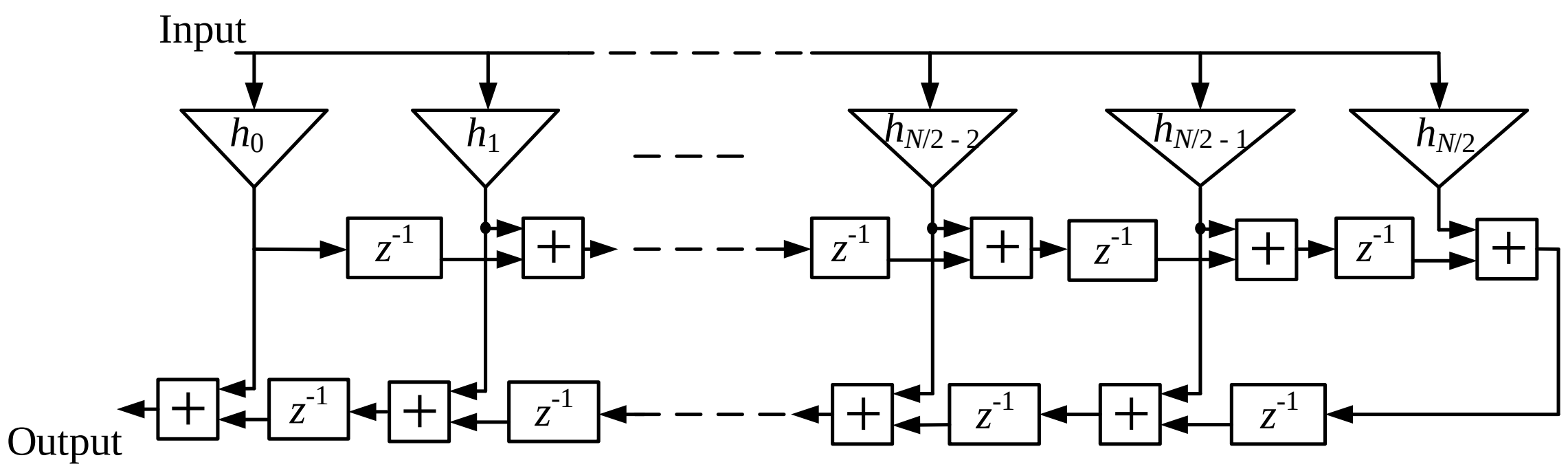

Fig. 2. Symmetric FIR filter implemented in transposed direct form structure.

## 2. Variable-Coefficient VCF Filters

The simplest method to obtain a variable FIR filter is to change all its coefficients as per the desired frequency response (i.e., in this case the various cutoff frequencies). All the corresponding filter coefficient sets are calculated in advance and are stored in a memory. For such a memory-based variable-coefficient (or reloadable or programmable) filter [4-10], shown in Fig. 3, appropriate coefficients are loaded onto the filter structure as per the desired cutoff frequency. When compared with Fig. 2, the coefficient multipliers here are variable multipliers (i.e. both the multiplicands are variable), and are represented by rectangles with '×'.

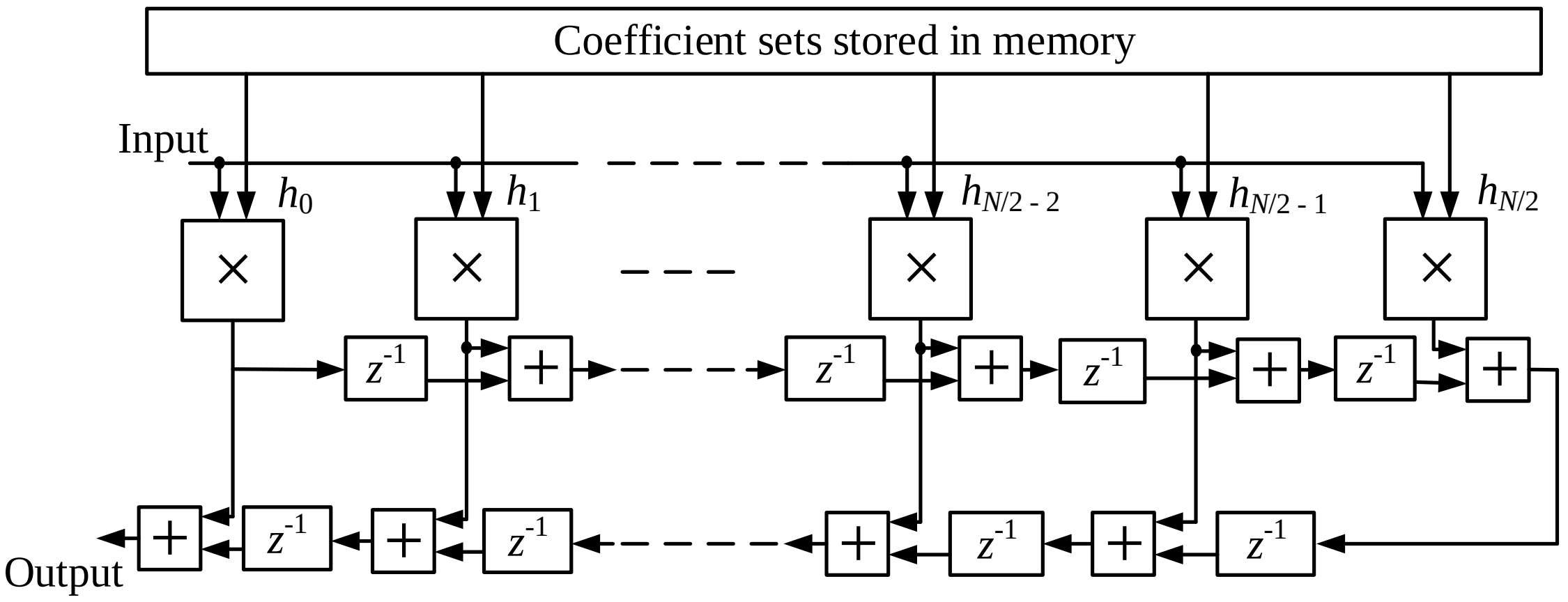

Fig. 3. Memory-based variable-coefficient filter.

Another approach to design variable-coefficient VCF filter is to calculate desired filter coefficients on-the-fly. A method in [11, 12] makes use of the well-known fact that, filter coefficients of an FIR filter can be expressed as linear (for center coefficient) and sinusoidal functions (for other coefficients) of its cutoff frequency. Based on these relations, first the coefficients of a prototype lowpass filter are calculated (using any standard method such as Remez exchange algorithm or the Parks-McClellan algorithm [55, 56]), and from these, the coefficients for other cutoff frequencies are approximated. Sine function values can be calculated using lookup table, series expansion, or a digital sinewave generator. This method is similar to the spectral parameter approximation (SPA) technique [38-51], main difference being the manner (on-line for [11, 12] and offline for SPA) in which the approximation is carried out. Also, as opposed to the SPA technique, it lacks clear and strong background, and results in less accurate VCF filter responses. Quadratic programming based method in [13] is a precursor of SPA technique. In this case, optimization procedure is performed offline, however, new coefficients corresponding to desired cutoff frequency need to be calculated on-the-fly using computationally heavy matrix multiplications. In [14], first a two-dimensional prototype filter is designed and its coefficients are stored. From these stored coefficients, filter coefficients for a 1-D filter are calculated for desired cutoff frequency. This procedure results in a more accurate frequency response than [11, 12], but has significantly higher computational cost for on-the-fly filter coefficient calculations. Due to on-the-fly calculation of filter coefficients, these methods may not be suitable for applications where high operating frequency is desired and cutoff frequency needs to be changed frequently.

Variable-coefficient filters are optimal in a sense that the filter order for the particular frequency response specification is minimum. The memory size required for a memory-based variable-coefficient filter is dependent on the filter order and the number of frequency responses to be obtained. Therefore, when the filter specifications are stringent, i.e., the filter order is relatively high, and the number of desired frequency responses is very large, the memory requirement for such a variable-coefficient filter is huge. In general, the complexity and time requirement for updating routines (i.e. reconfiguration time) increases with the increase in the filter order, due to a large number of memory access operations and on-the-fly calculation of filter coefficients for [11-14]. Thus, variable-coefficient filters are appropriate for realizing only lower order filters. In addition, a memory-based variable-coefficient filter is not field-upgradable in the sense that it can cater to only those frequency responses for which the filter coefficients are already stored in the memory. All these limitations make the variable-coefficient filters unsuitable for the applications where the stringent specifications of the desired responses change dynamically and frequently. However, major advantage of memory-based variable-coefficient filters is that they can be used for obtaining (pre-calculated) frequency responses which are arbitrary and have no relation whatsoever. As opposed to this, when the desired variable frequency responses have some similarities or are related to each other (e.g. when desired cutoff frequencies are multiples of one particular cutoff frequency), generally, fixed-coefficient VCF filters [15-54] provide area-efficient alternatives.

## 3. Fixed-Coefficient VCF Filters

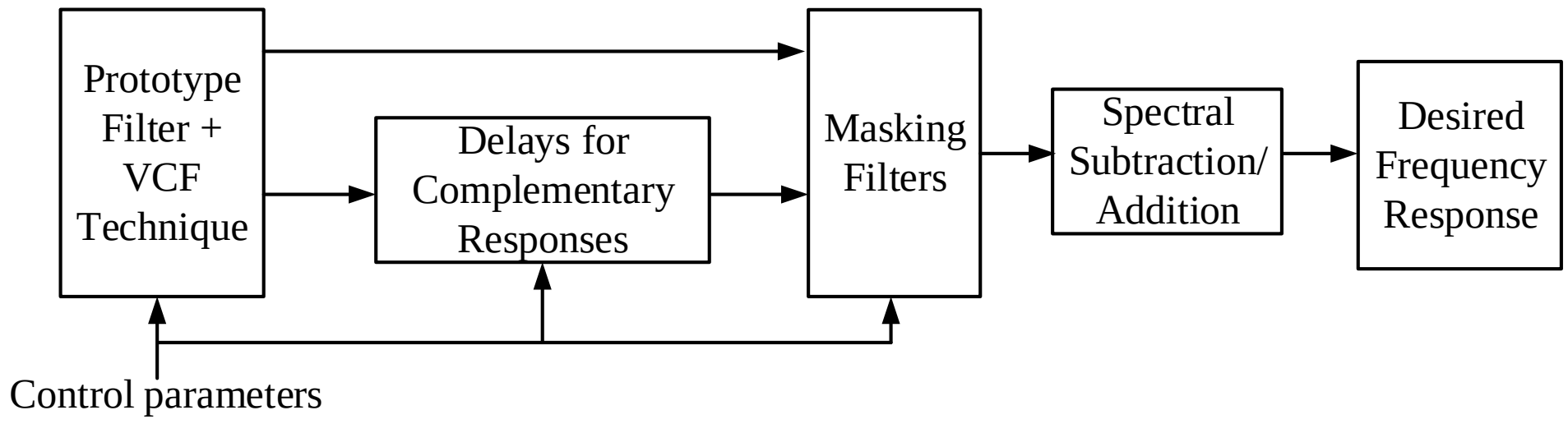

Fig. 4. Generalized block diagram for VCF filter design.

Different types of fixed-coefficient VCF filters are discussed in detail in this section. Fig. 4 presents a generalized block diagram for designing a fixed-coefficient VCF filter. The corresponding design procedure is given below -

*Step-1:* For the application under consideration in which a VCF filter is to be employed, identify the desired cutoff frequency control - discrete or continuous and its range, allowable variations of frequency response parameters ($\{tbw, \delta_p, \delta_s\}_{_constraints}$, group delay, and phase) and the complexity constraints.

*Step-2:* Based on the qualitative and quantitative comparisons of different VCF filters provided in Sections 3.1 (VCF filters with discrete cutoff frequency control) and 3.2 (VCF filters with continuous cutoff frequency control) respectively, choose the most suitable VCF filter design technique.

*Step-3:* Identify the required values of the control parameters. Determine the prototype filter's specifications $f_{c_proto}$ and $\{tbw, \delta_p, \delta_s\}_{_proto}$ based on $\{tbw, \delta_p, \delta_s\}_{_constraints}$ and the chosen VCF filter design technique.

*Step-4:* Obtain the coefficients for prototype filter (and masking filters if required) using a suitable FIR filter design technique.

*Step-5:* Implement the overall VCF filter using the corresponding filter implementation architecture to obtain the desired frequency responses by varying the controlling parameters. If required, spectral subtraction/addition operations and complementary frequency responses can also be employed to obtain the desired frequency responses. Complementary filter response can be obtained by subtracting its output from a suitably delayed version of input.

The prototype filter (usually a lowpass filter) can be designed using any standard method such as [55, 56]. The VCF filter realized using this prototype filter is also a variable lowpass filter. Once a VCF lowpass filter is obtained, a VCF highpass or bandpass or bandstop filter can be realized with trivial modifications, and hence these are not discussed separately in detail.

A desirable quality for a VCF filter is minimum overhead on the area and power consumption, along with its cutoff frequency being controllable through small number of parameters (to make its reconfiguration easier). Also, small number of variable parameters means that fewer variable blocks in the filter architecture, and consequently, more fixed blocks. Once designed, the fixed coefficients in a VCF filter architecture can be implemented using hardware reduction techniques such as MCM blocks [69, 70]. In general, one can incorporate optimization techniques (e.g. [71]) or adjust the ranges of the parameters (e.g. [46]) in Step-3 or Step-4 in order to obtain the prototype and masking filter coefficients with desired characteristics. Such optimizations to reduce the bit-widths of coefficients, area of the overall filter can be used on all of the fixed-coefficient VCF design techniques discussed in this section. Even though not covered in this paper, we recommend a joint optimization approach for designing area- and power-efficient VCF filters. It is to be noted that variable multipliers usually have larger area and delay and consume more power compared to fixed multipliers and such area reduction techniques cannot be directly applied to variable-coefficient VCF filters.

Based on the degree of control over the cutoff frequency they offer, the VCF filters can be categorized as the VCF filters with *discrete* control over the cutoff frequency (variable-coefficient filters [4-14], coefficient decimation based filters [15-19], frequency response masking based filters [20-25]) and the VCF filters with *continuous* control over the cutoff frequency (all-pass transformation based filters [26-29], fractional delay structure based filters [30-32], frequency transformation based filters [33-37], SPA based filters [38-54]). Here discrete control means only a limited number of cutoff frequencies can be obtained using that particular technique. For e.g., if a prototype filter with cutoff frequency 0.12 is used along with the coefficient decimation II technique (explained in Section 3.1.1), maximum of 8 cutoff frequencies, i.e., {0.12, 0.24, 0.36, 0.48, 0.60, 0.72, 0.84, 0.96} can be obtained.

Continuous (unabridged) control refers to the ability of the technique to have very fine cutoff frequency resolution of the order of 0.001 or higher, which is usually restricted only by the allotted number of bits for the controlling parameter in the corresponding fixed-point implementations. Variable-coefficient filters, discussed in Section 2, can be designed for such high resolutions, but they require prohibitively large memory to store all the coefficient sets.

In general, the transition bandwidth of the VCF filter varies according to the parameter values and transition bandwidth of prototype filter. In this paper, a parameter $\psi$ (ratio of transition bandwidth of VCF filter to transition bandwidth of prototype filter) is used to illustrate this relation. Higher the value of $\psi$, wider is the bandwidth of the variable filter response. Thus its maximum value ($\psi_{max}$) is an important criterion in any VCF filter design technique to determine the transition bandwidth of the prototype filter, so that the transition bandwidth of the VCF filter is within its final desired specifications.

## 3.1 VCF Filters with Discrete Control over Cutoff Frequency

### 3.1.1 Coefficient Decimation based VCF Filters

Coefficient decimation method (CDM) involves the selective usage of filter coefficients by performing operations such as replacing them by zeros and retaining/discarding them appropriately to obtain VCF filter responses [15, 16]. CDM consists of two coefficient decimation operations - CDM-I and CDM-II [16]. In the CDM-I operation, coefficients of a lowpass prototype (original) filter are decimated by a factor $D$, i.e., every $D^{th}$ coefficient is retained and the others are replaced by zeros, to obtain a multi-band frequency response. The resultant frequency response has its subbands located at center frequencies given by even multiples of $1/D$, i.e., $2k/D$ where $k$ is an integer ranging from 0 to ($D$-1). In the CDM-II operation, every $D^{th}$ coefficient is retained and all others are discarded to obtain a lowpass frequency response with its cutoff frequency and transition bandwidth

*D* times that of the prototype filter. The resultant frequency responses after performing CDM operations can be subjected to complementary filter operation and appropriate arithmetic operations to obtain the desired subbands, followed by use of masking filters if needed.

The frequency responses of the prototype filter, and the coefficient decimated filters after CDM-I by $D = 2$ and 4 are shown in Fig. 5. The frequency responses of the prototype filter, and the coefficient decimated filters after CDM-II by $D = 2$ and 4 are shown in Fig. 6. It can be noted that the passband ripple (though not visible in these figures) and stopband attenuation has deteriorated for CDM-I and CDM-II, and the transition bandwidth of the coefficient decimated filter has increased by a factor of *D* for CDM-II.

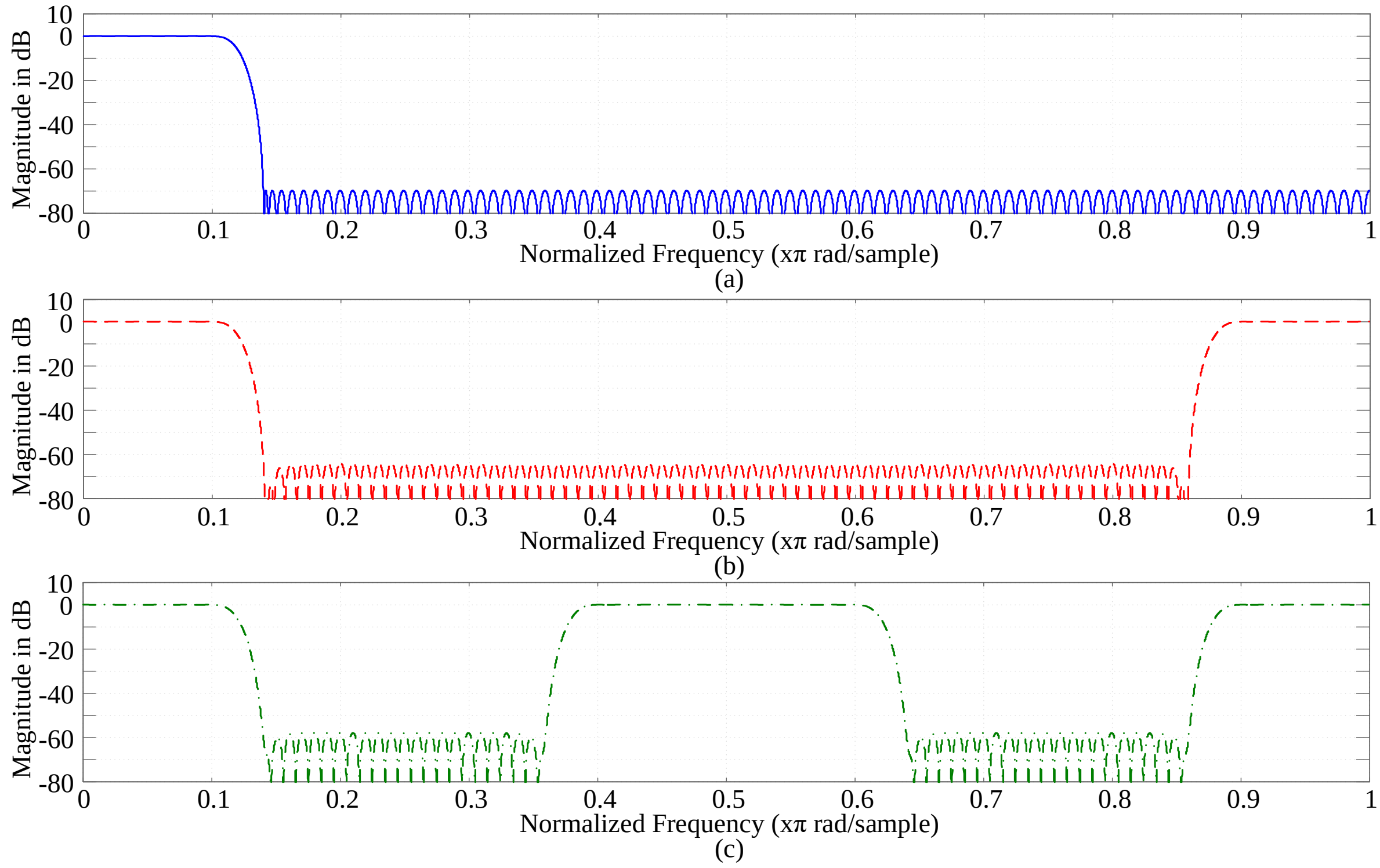

Fig. 5. Magnitude responses of (a) prototype filter, (b) filter after CDM-I by $D = 2$ on prototype filter, (c) filter after CDM-I by $D = 4$ on prototype filter.

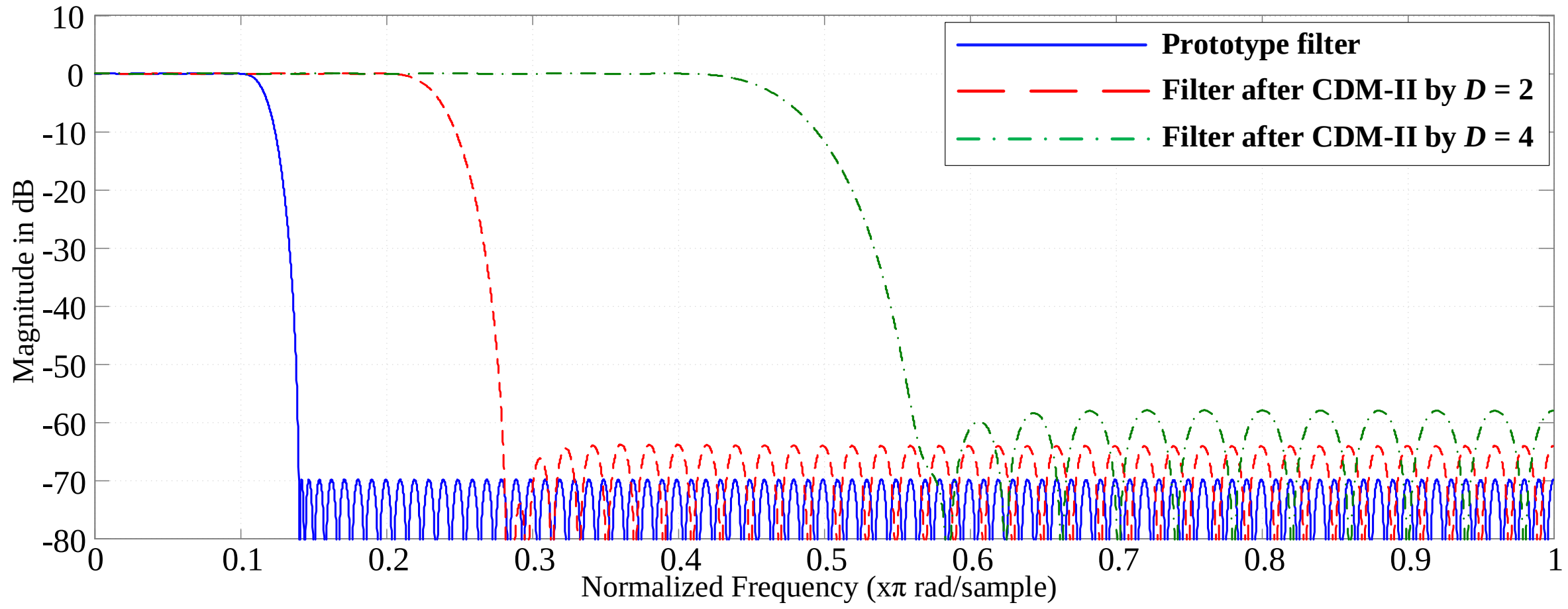

Fig. 6. Magnitude responses of prototype filter and filters after CDM-II by $D = 2$ and 4 on prototype filter.

A modified coefficient decimation method (MCDM) was proposed in [17]. Similar to CDM, MCDM consists of two coefficient decimation operations - MCDM-I and MCDM-II [18]. In the MCDM-I operation, if the prototype filter is decimated by a factor $D$, every $D^{th}$ coefficient is retained and the sign of every alternate retained coefficient is reversed. All other coefficients are replaced by zeros. This results in a multi-band frequency response with center frequency locations of the subbands given by odd multiples of $1/D$, i.e., $(2k+1)/D$ where $k$ is an integer ranging from 0 to ($D$-1). In the MCDM-II operation, every $D^{th}$ coefficient is retained and all others are discarded. The sign of every alternate retained coefficient is reversed to obtain a highpass frequency response with its bandwidth $D$ times that of the prototype filter.

The combination of CDM and MCDM is termed as improved coefficient decimation method (ICDM), which consists of four distinct coefficient decimation operations – CDM-I, CDM-II, MCDM-I and MCDM-II. These operations are classified as ICDM-I (includes CDM-I and MCDM-I) and ICDM-II (includes CDM-II and MCDM-II) [18]. It can be noted that center frequency resolution of $1/D$ can be achieved for the subbands in the output frequency responses obtained after performing ICDM-I operations. On the other hand, by performing ICDM-II operations, subband bandwidths that are integer multiples of that of the prototype filter can be achieved by using appropriate values of $D$.

Fig. 7 shows a consolidated hardware implementation architecture which can be used to implement coefficient decimation based VCF filters. Multiplexers form the most important components of this architecture and are used to retain/discard appropriate coefficients. Multiplexers labeled 'mux I' are used while performing CDM-I and MCDM-I operations while those labeled 'mux II' are used while performing CDM-II and MCDM-II operations. The 'add/sub' blocks are used to perform sign reversal of filter coefficients in MCDM operations. The select lines of the multiplexers and the add/sub blocks are triggered by the decimation selector logic. The variable multiplier $D_{out}$ is used to scale the resulting output to obtain the original magnitude response. Details of specific hardware implementation architectures for the different coefficient decimation techniques are available in [15-18].

A major drawback of coefficient decimation is that passband ripple and stopband attenuation in the resultant frequency responses deteriorates (by factor of $D$) as the value of $D$ is increased. For CDM-II and ICDM-II, $\psi \geq 1$, and $\psi_{\max} = D_{\max}$, where $D_{\max}$ is the maximum value of coefficient decimation factor considered in the design procedure. To address these problems, overdesign of the prototype filter is used which results in the requirement of higher order prototype filters that increase the implementation complexity of the VCF filters. Linear programming based filter design technique proposed in [19] can also be used to address this problem. In [19], prototype filter design is considered as an optimization problem with the required set of coefficient decimation factors and desired stopband attenuations considered beforehand as constraints. However, this approach results in different filter coefficients for each set of decimation factors. Hence, it is not a feasible approach and advanced optimization techniques need to be explored.

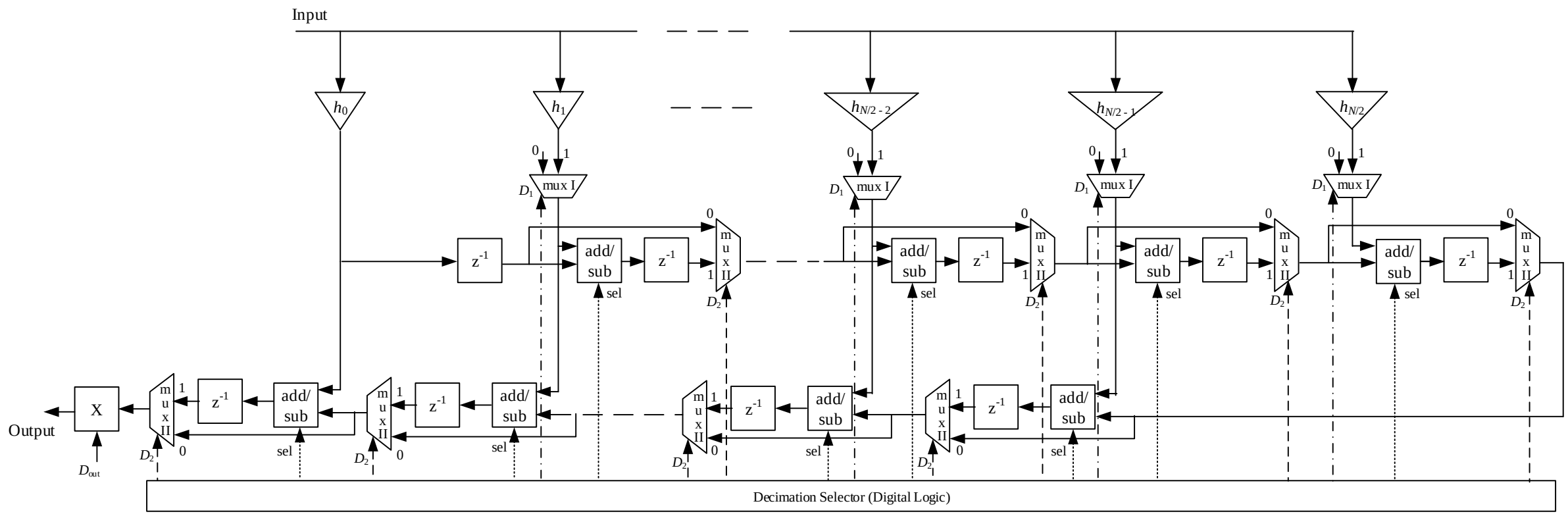

Fig. 7. Hardware implementation architecture for coefficient decimation based VCF filter.

### 3.1.2 Interpolation - Frequency Response Masking based VCF Filters

The interpolation approach was first proposed to obtain a low complexity narrow transition bandwidth filter from a relatively wider transition bandwidth filter [57-59] and was later adapted for the design of VCF filters [20, 21]. In interpolation or frequency response masking (FRM), $M$-1 zeros are inserted between the successive coefficients the prototype filter, and then this new impulse response is interpolated by using a suitable cascaded filter. Insertion of zero-valued coefficients results in a multiband response, replicating the prototype filter's response at the multiples of $2/M$. The bandwidths and transition bandwidths of these bands are $1/M$ times those of the prototype filter. The complementary response of this interpolated filter response is also obtained simultaneously. Cascaded filters are used to extract the desired band(s) and masking other bands, and hence are termed as 'masking filters'. Appropriate bands from both the multiband responses are extracted and added together to obtain a narrow transition bandwidth filter (with $\psi = 1/M$). The prototype filter response, the interpolated multiband response and its complementary response for $M = 4$, and the variable cutoff frequency responses obtained by addition of different number of bands from these two responses are shown in Fig. 8. VCF filters are obtained by a) varying the number of bands to be extracted and added for one particular value of $M$ and/or b) by varying the value of $M$.

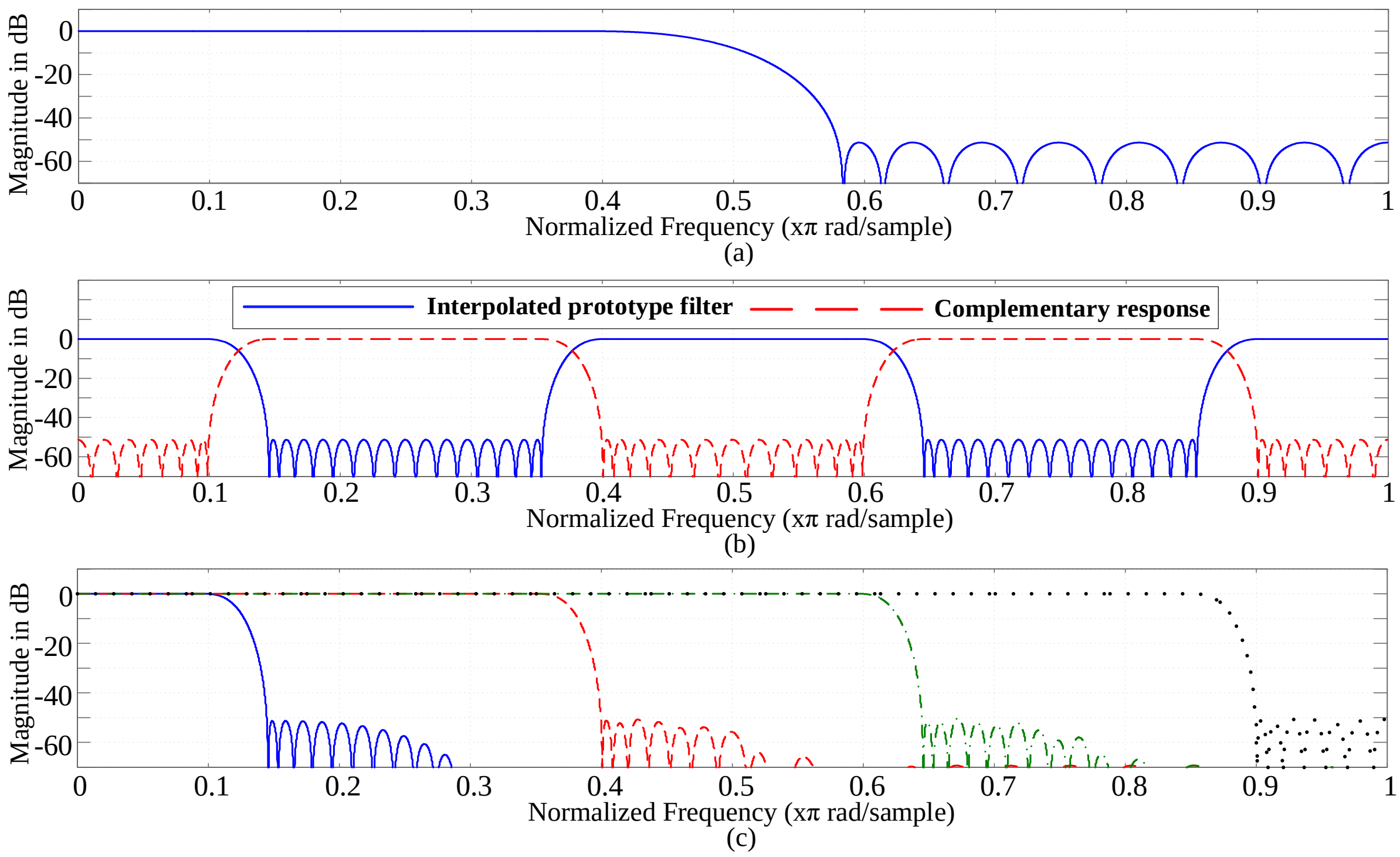

Fig. 8. Magnitude responses of (a) prototype filter, (b) interpolated prototype filter and its complementary response, (c) various lowpass responses obtained by addition of adjacent extracted bands from the responses shown in (b).

Prototype filter of the interpolation based VCF filter is shown in Fig. 9. Insertion of $M$-1 zeros between two coefficients of the prototype filter is accomplished by replacing a unit delay in the add-delay chain of the filter structure by $M$ delays (shown in Fig. 9 as $z^{-M}$). Reconfiguration in terms of the interpolation factor is achieved using a multiplexer, where a common select line is used to select appropriate number of delays at every multiplexer. Select line is not shown in Fig. 9 for maintaining the clarity. The complementary response of the interpolated prototype filter is obtained by subtracting its output from the suitably delayed input signal. The prototype filter response and complementary response are processed by fixed-coefficient masking filter(s) for extracting and adding desired bands.

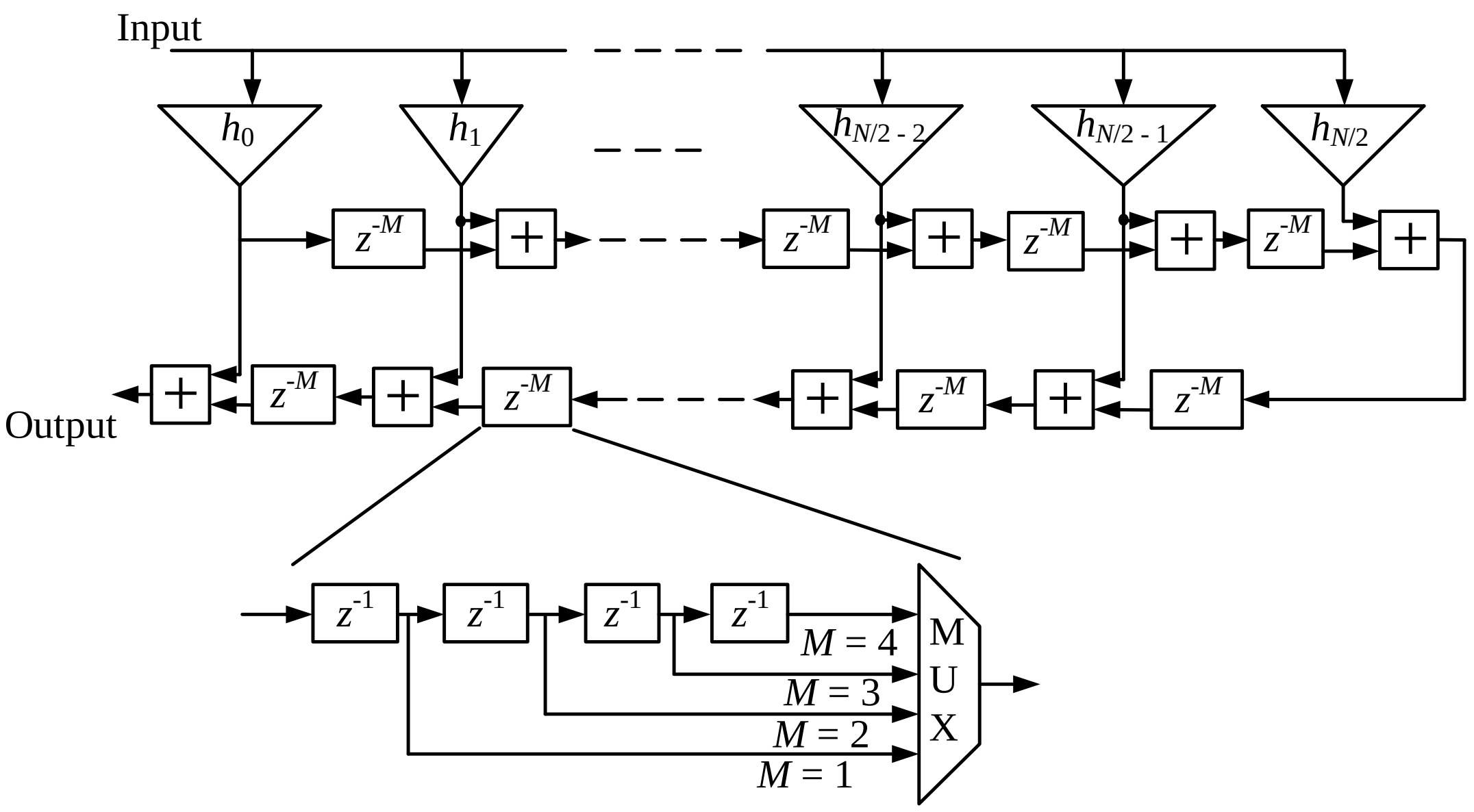

Fig. 9. Prototype filter of interpolation based VCF filter.

Therefore, using only the fixed-coefficient filters, various lowpass, highpass, bandpass, and bandstop responses can be obtained from the FRM based filter, without any hardware re-implementation. Note that $\psi \leq 1$ and $\psi_{\max} = 1/M_{\min}$, where $M_{\min}$ is the minimum value of interpolation factors considered in the design procedure.

The classic two-stage FRM approach mentioned above is extended in [22, 23] to a multi-stage design, in which a low complexity variable filter based on the fast filter bank is proposed. In the fast filter bank, from second stage onwards, interpolated and frequency shifted masking filters are used to extract bands from the interpolated response of the prototype filter in the first stage and its complementary response. In [22, 23], appropriate number of subbands in the fast filter bank are selected, and the last subband from this selection is shaped using a shaping filter to obtain very narrow transition bandwidth with fine (but discrete) control over the cutoff frequency. However, this fast filter bank based filter has a very large group delay as it uses the fast filter bank as well as a multistage shaping filter. A detailed discussion on design and implementation of fast filter bank can be found in [60, 61].

In general, FRM based filters have large group delay. Also, it becomes difficult to choose appropriate values of $M$, when desired cutoff frequencies are closely spaced and do not have simple relations with each other. In order to overcome this drawback, the coefficient decimation and interpolation techniques can be combined to design the variable filter with better control over cutoff frequency [24]. Its enhanced version, 'Improved Coefficient Decimation-Interpolation-Masking (IDIM)' method in [25], provides center frequency resolution of $1/M$ (compared to $2/M$ in [24]) and approximately twice the number of filter responses for same design parameters. In these techniques, the bandwidth of the prototype filter is first increased by applying CDM-II operation (factor $D$) on the prototype filter, and multiband response with narrower bandwidth is then obtained by using interpolation (factor $M$). In [25], additionally, lowpass-to-highpass transformation is used. This transformation consists of reversing the sign of every alternate coefficient of the filter, whose highpass version is to be obtained. It is implemented using 'add/sub' blocks and a select line. Appropriate bands are then extracted from the interpolated filter response and its complementary response to obtain variable bandwidth lowpass responses, with bandwidths that are multiples of ($D/M$ × prototype filter's bandwidth), with $\psi_{\max} = D_{\max}/M_{\min}$.

A qualitative comparison of the above mentioned techniques is provided in Table I. We note that all these techniques result in linear-phase VCF responses when the prototype filter has linear phase. Except for coefficient decimation type II techniques, using the same lowpass prototype filter one can obtain bandpass responses directly by varying $D$ or $M$. Another add-delay chain can be implemented for coefficient decimation type II filters, in order to obtain second lowpass response (with different value of $D$), and a bandpass response can be obtained from two simultaneously obtained lowpass responses. Due to simple architecture, all these techniques are suitable for designing a filter bank. The center frequency resolution for bandpass responses (and that of the subbands in the filter bank) depends on the corresponding coefficient decimation or interpolation factor. When multiple values of $D$ and/or $M$ are considered, it depends on the maximum value of $D$ or $M$ as mentioned in the table.

Table I: Qualitative comparison of VCF filters with *discrete* control over cutoff frequency.

| | Cutoff frequency range | *tbw* of VCF responses | Flexibility [a] | Implementation complexity [b] | Group delay [c] | Suitable for narrow *tbw* | Center frequency resolution for bandpass responses |
|---|---|---|---|---|---|---|---|
| Coefficient decimation - type I (CDM-I, MCDM-I, ICDM-I) | Nyquist band | Fixed | Medium | Medium | $> N/2$ | Yes | $2/D_{max}$ for CDM, $1/D_{max}$ for ICDM |
| Coefficient decimation - type II (CDM-II, MCDM-II, ICDM-II) | Nyquist band | Variable | Low | High | $> N/2$ | No | |
| Interpolation - Frequency response masking | Nyquist band | Variable | Low | Very low | $>> N/2$ | Yes | $2/M_{max}$ |
| Coefficient decimation + interpolation | Nyquist band | Variable | High | Low | $> N/2$ | Yes | $2/M_{max}$ |
| Improved coefficient decimation + interpolation | Nyquist band | Variable | Very high | Low | $> N/2$ | Yes | $1/M_{max}$ |

[a] Relative flexibility for techniques considered in Table I is in terms of number of distinct frequency responses which can be obtained using a single set of prototype filter coefficients.

[b] Relative complexity for techniques considered in this table for same/similar $\{tbw, \delta_p, \delta_s\}_{_constraints}$ while obtaining a desired frequency response. By complexity, we refer to number of multipliers required to realize the filter.

[c] $N$ denotes the order of the variable-coefficient filter designed to obtain same cutoff frequencies with same $\{tbw, \delta_p, \delta_s\}_{_constraints}$. Group delay of this filter is $N/2$.

In Table II we provide a quantitative comparison for the VCF filters designed with $f_{c_proto} = 0.085$ and $tbw_{_proto} = 0.03$, $D = \{1, 2, 3, 4\}$ and $M = \{1, 2, 3, 4, 5, 6, 7, 8\}$. Appropriate masking filters are designed to extract variable lowpass, bandpass, and highpass responses. Table II provides the comparison of these filters in terms of total number of multipliers required to realize a particular VCF filter considering symmetric coefficients for prototype and masking filters. Number of multiplications refers to the maximum number of multiplications required in order to obtain one filter response, and it is less than or equal to the total number of multipliers. The third column of Table II provides the total number of distinct responses provided by that particular variable filter. The coefficient decimation based filter and frequency response masking filter are designed for same specifications as in the case of narrowest band obtained in IDIM based filter. Coefficient decimation factors of up to $D = 9$ can be used for CDM-II based filter, however due to narrow transition bandwidth specifications its complexity (total number of multipliers required) will be even higher. As the center frequency resolution of ICDM (1/*D*) is half that of CDM (2/*D*), an ICDM based filter can provide approximately twice the number of filter responses at similar filter complexity. It is to be noted that when transition bandwidth specifications are relaxed, the coefficient decimation techniques can provide similar frequency response flexibility at lower complexities similar to the other techniques.

Table II: Quantitative comparison of VCF filters with *discrete* control over cutoff frequency.

| | Number of multipliers | Number of multiplications | Number of responses obtained |
|---|---|---|---|
| CDM based filter [16] | 901 | 901 | 46 |
| FRM based filter [58] | 123 | 123 | < 46 |
| DIM based filter [24] | 122 | 122 | 96 |
| IDIM based filter [25] | 261 | 122 | 176 |

Referring to Table I, even though the cutoff frequency values are spread out in the entire Nyquist band (0 to 1), all these VCF filters [15-25] can provide only a discrete control over the cutoff frequency, as the coefficient decimation and interpolation factors can assume only positive integer values. Therefore, these VCF filters are not field-upgradable in the sense that, the number of distinct cutoff frequencies that can be obtained is dependent on the values of *D* and *M* that are considered at the time of filter design and implementation, and the cutoff frequency values other than these values are not possible. When the desired cutoff frequencies are not integer or fractional multiples of each other, or a very high resolution is desired in a particular frequency range, these VCF filters are not suitable. Following example illustrates this point.

*Example*: Consider the desired cutoff frequencies to be 0.27, 0.30 and 0.33 with desired $tbw \leq 0.1$. In such a case, for CDM-II based VCF, prototype filter with $f_c = 0.03$ and $tbw \leq 0.009$ is required with $D = 9$, 10 and 11. However, if the desired cutoff frequencies are changed slightly, e.g. as 0.27, 0.29 and 0.33, prototype filter should be designed with $f_c = 0.01$ and $D = 27$, 29 and 33. Clearly, CDM is not a suitable technique for such designs. Same holds for FRM and the combinations of CDM and FRM.

In the case of cognitive radio networks, different users may use multiple wireless communication standards. These standards may have distinct bandwidths without any relation between them (e.g. integer of fractional multiplicity) [68]. In current LTE standard, the transmission bandwidth is limited to 1.4 MHz, 3 MHz, 5 MHz, 10 MHz, 15 MHz and 30 MHz. In next generation systems, we may have finer control over the bandwidth and allowable bandwidth can be any value which is an integer multiple of the bandwidth of the resource block (which 180 KHz in existing standard). In such cases, it is desirable to have a much more flexible control over cutoff frequency (a continuous control, cutoff frequency resolution of 0.001 or higher) than offered by the techniques we have discussed till now. (It can be noted that the fast filter bank based filter [22, 23] can be used in the above Example. However, its complexity and group delay increase significantly when higher resolution and wider cutoff frequency range is desired.) Next, we will consider the VCF filters with continuous control over the cutoff frequency. In order to illustrate the fine resolution that is provided by such filters, Fig. 10 shows total 51 VCF filter responses obtained from second-order frequency transformation based filter. The cutoff frequency resolution is approximately 0.0053. Similar filter responses are obtained using other design techniques that are discussed in the next section.

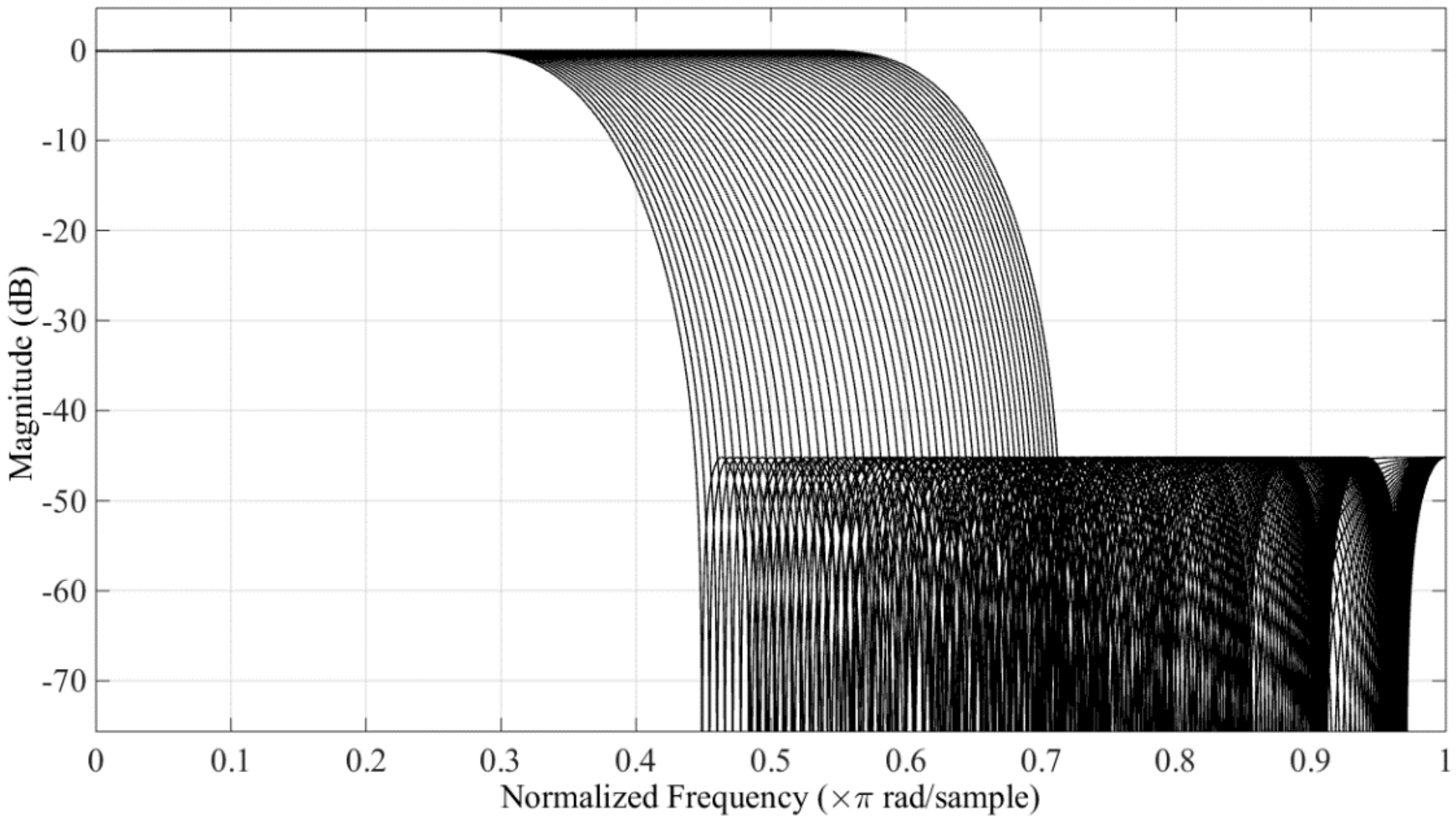

Fig. 10. VCF filter responses obtained from second-order frequency transformation based filter (cutoff frequency resolution of 0.0053).

## 3.2 VCF Filters with Continuous Control over Cutoff Frequency

### 3.2.1 All-Pass Transformation based VCF Filters

The all-pass transformation **(**APT) based VCF filter design technique was first proposed in [26]. In the APT based VCF filter, every unit delay element in the prototype filter is replaced by an all-pass filter structure of an appropriate order [62]. The all-pass filter coefficients are then varied on-the-fly to achieve the desired frequency responses with variable cutoff frequencies. APT based VCF filters are also known as warped filters as the output frequency responses are warped versions of the prototype filter response. APT based VCF filters offer continuous control on the cutoff frequency over the entire Nyquist frequency range. They involve lesser design overheads and achieve lower implementation complexities when compared with the frequency transformation based VCF filters and SPA based VCF filters. However, APT based VCF filter is a non-linear-phase filter even if the prototype filter is a linear-phase filter. Also, obtaining complementary responses is not straightforward in APT based VCF filters. This makes the design of reconfigurable filter banks computationally complex compared to other VCFs.

Hardware implementation architecture of the APT based VCF filter is shown in Fig. 11(a), where all-pass filter structure is represented by $P(z)$. Two types of APT based VCF filters are typically used in various applications – first-order APT based and second-order APT based. In a first-order APT based VCF filter, every delay element in prototype filter structure is replaced by a first-order all-pass filter structure (shown in Fig. 11(b)), to obtain variable lowpass and highpass frequency responses as a function of a single tuning parameter $\alpha$. Similarly, second-order APT based VCF filter employs a second-order all-pass filter structure (shown in Fig. 11(c)), and therefore uses two tunable parameters ($\beta_2$ and $\beta_3$) which can be separately controlled. In turn, variable lowpass, highpass as well as bandpass and bandstop frequency responses with independent and continuous control over both their cutoff frequencies can be obtained. This flexibility comes at the cost of significantly higher complexity than the first-order APT based VCF filter, as usage of second-order all-pass filter structure leads to requirement of a greater number of variable multipliers. The mathematical relation between the warping coefficients and the cutoff frequencies for both first- as well as second-order APT based VCF filters is available in [26, 62]. For any prototype filter, $\psi_{\max}$ is observed when cutoff frequency of the APT based VCF filter is 0.5, and in general, $\psi_{max} \geq 1$. In [27], modified pipelined hardware implementation architectures for first- and second-order APT based VCF filters are presented to realize them with high operating frequencies that are independent of the prototype filter order.

To enable the realization of variable lowpass, highpass, bandpass and bandstop frequency responses at a lower complexity than the second-order APT based VCF filter, designs combining all-pass transformation with coefficient decimation techniques were recently proposed in [28, 29]. VCF filter based on the combination of CDM [15] and first-order APT based technique was proposed in [28] and a lower complexity alternative to it was proposed in [29] by combining ICDM [18] and the first-order APT. Hardware implementation architectures and the corresponding mathematical formulations for these modified techniques are available in [28, 29].

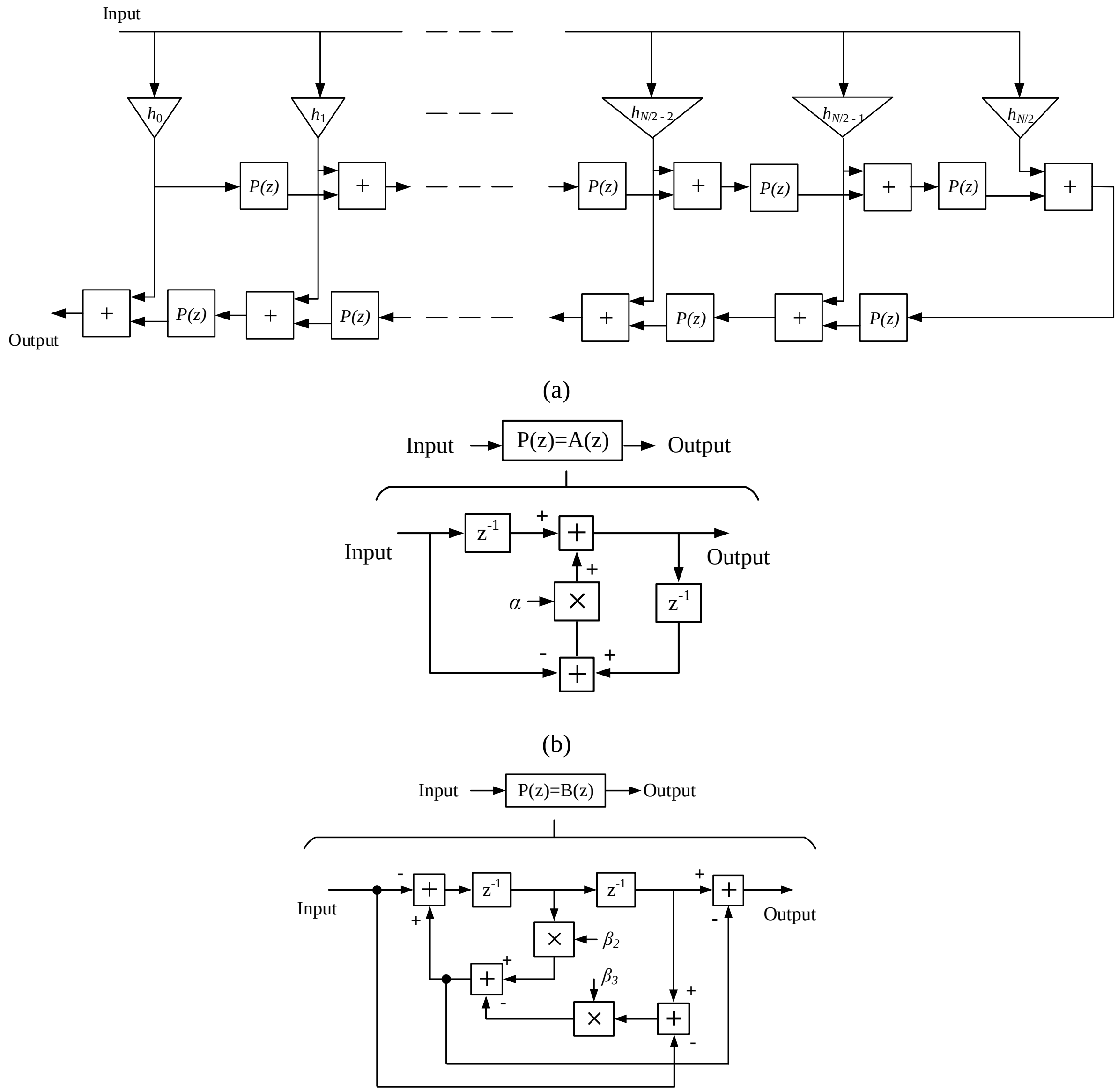

(c)

Fig. 11. Hardware implementation architecture of (a) APT based VCF filter, (b) first-order all-pass structure, (c) second-order all-pass structure.

### 3.2.2 Fractional Delay Structure based VCF Filters

In case of interpolation based filters, every unit delay of the prototype filter is replaced by $M$ unit delays and such an interpolation results in decreasing the cutoff frequency by factor of $1/M$. As the value of $M$ can only be an integer greater than one (discrete values), the cutoff frequencies obtained by interpolation and masking technique are discrete, i.e., only limited set of values can be obtained.

As the cutoff frequency is inversely proportional to interpolation factor, it can be varied *continuously* by making interpolation factor to vary *continuously*. This means that the unit delay of the filter structure needs to be replaced by a fractional delay. (A detailed literature review regarding the 'fractional delay structures' (FDSs), which can provide desired fractional delay, can be found in [1]. More recent advances in the design of FDSs are [63-66].) To control the cutoff frequency of the overall VCF filter on-the-fly, the fractional delay generated by the FDS should be variable on-the-fly. Moreover, such a variable FDS should be of lower order and have low complexity, so as to maintain low complexity for overall VCF filter.

An FDS in [67], designed using second-order modified Farrow structure, satisfies all the above mentioned requirements. Being a second-order FDS, it provides a fractional delay equal to $1+d$, where $d$ is the tuneable parameter in the range $0 \leq d < 1$. Using this FDS from [67], an FDS based VCF filter is proposed in [30]. Cutoff frequency and transition bandwidth of the FDS based VCF filter become $1/(1+d)$ times those of the prototype filter. Cutoff frequency of the FDS based VCF filter ($f_c$) can be varied in the range $f_{c_proto}/2 < f_c \leq f_{c_proto}$, where $f_{c_proto}$ is the cutoff frequency of the prototype filter.

The second-order modified Farrow structure can provide unity magnitude response and constant phase response only for low frequencies (approximately up to 0.2) [1, 67]. Therefore, the frequency response of the FDS based VCF filter deteriorates for higher cutoff frequencies, and the maximum cutoff frequency obtained from the FDS based VCF filter can be approximately 0.2. Assuming the cutoff frequency of the prototype filter to be 0.2, the cutoff frequency of the FDS based VCF filter can be varied only in the limited range as $0.1 < f_c \leq 0.2$.

In [30], CDM-II is used to increase the cutoff frequency range of the FDS based VCF filter. For $f_{c_proto} = 0.1$, the cutoff frequency of the FDS based VCF filter varies as $0.05 < f_c \leq 0.1$. By using CDM-II on this filter, the cutoff frequencies for the range $0.1 < f_c \leq 0.2$ can be obtained. However, the prototype filter needs to be overdesigned, i.e., its order should be increased, in order to compensate for the $\delta_p$, $\delta_s$ and $tbw$ degradation ($\psi = D_{\max}/(1+d)$) which is inherent to the CDM-II technique.

An intuitive way to increase the cutoff frequency range of the FDS based VCF filter is to increase the range of the fractional delay. Larger fractional delay range can be obtained by using the higher-order FDS. However, the use of higher-order FDS means that more number of multiplications per FDS which will be used to replace the unit delays in the prototype filter. Therefore, the use of higher-order FDS will increase the number of multiplications required in the overall FDS based VCF filter.

In [31], a continuously variable fractional delay (CVFD) element is proposed which is derived from the modified Farrow structure based fractional delay element from [67]. When compared with existing FDSs [1, 63-67], the CVFD element provides wide fractional delay range at the minimum multiplication complexity possible, and is capable of changing the fractional delay range on-the-fly (i.e. from $1 \leq 1+d < 2$ to $2 \leq 2+d < 3$ etc.). Use of this CVFD element to design an FDS based VCF filter increases the cutoff frequency range below its lower limit of $f_{c_proto}/2$.

To further increase the cutoff frequency range and resolution, interpolation technique is incorporated in these filter designs in [31, 32]. In [32], the interpolated and complementary responses are used together to obtain VCF lowpass responses in the Nyquist band. For instance, the FDS based VCF filter in [32] has cutoff frequency range 0.013 to 0.987. It provides coarse as well as fine control over the cutoff frequency. However, the control over the cutoff frequency is still in disjoint ranges.

Generic architecture of an FDS based VCF filters [30-32] is shown in Fig. 12. Parameters $d$, $M$ and $p$ control the variable fractional delay, and therefore the cutoff frequency of the overall VCF filter. Wherever required, suitable masking filter needs to be cascaded in order to extract the desired lowpass/bandpass/highpass responses. Note that $\psi = 1/(p + (1+d)\times M)$. Generic architecture of the FDSs used in [31, 32] is shown in Fig. 13. For $p = 0$ and $M = 1$, it is equivalent to the modified second-order Farrow structure used in [30].

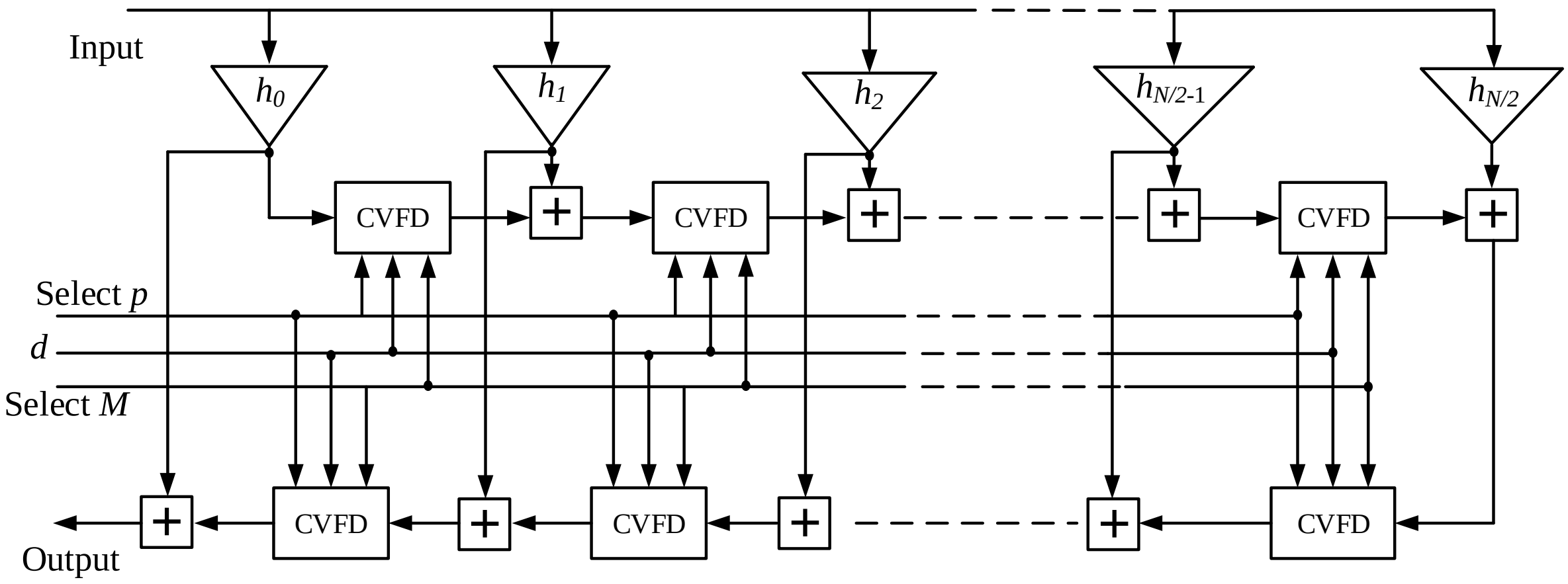

Fig. 12. Architecture of FDS based VCF filter. Parameters $d$, $M$ and $p$ control the fractional delay, and therefore, the cutoff frequency of the overall VCF filter.

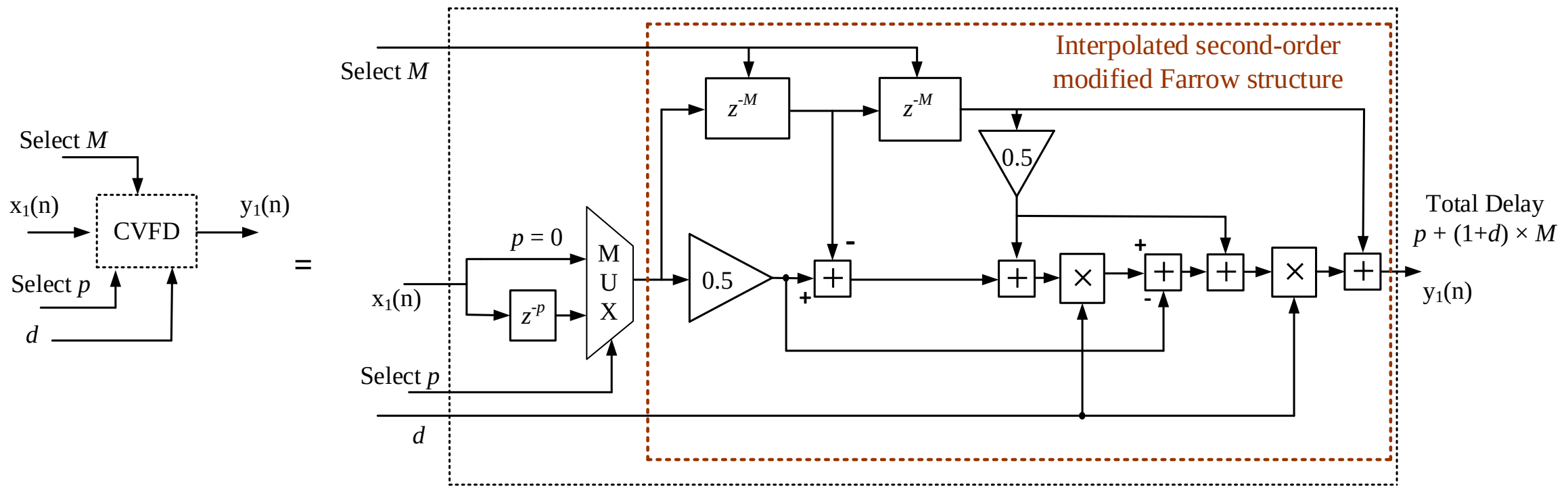

Fig. 13. Architecture of FDS used in [31, 32]. It reduces to the modified second-order FDS used in [30] for $p = 0$ and $M = 1$.

### 3.2.3 Frequency Transformation based VCF Filters

The frequency transformations are one of the first techniques explored for design of VCF filter [33, 34]. A variable impulse response filter is obtained by applying frequency transformation on the Taylor expansion of the impulse response of the prototype filter. Fig. 14 shows the hardware implementation architecture for a $P^{th}$-order frequency transformation based VCF filter for a prototype filter of order $N$. The coefficients $a_n$ of this Taylor structure are related to the symmetric coefficients $h_n$ of the prototype filter via the Chebyshev polynomials. The coefficients $a_n$ have fixed values for a particular prototype filter, and the cutoff frequency of the frequency transformed filter is controlled by varying the coefficients $A_i$.

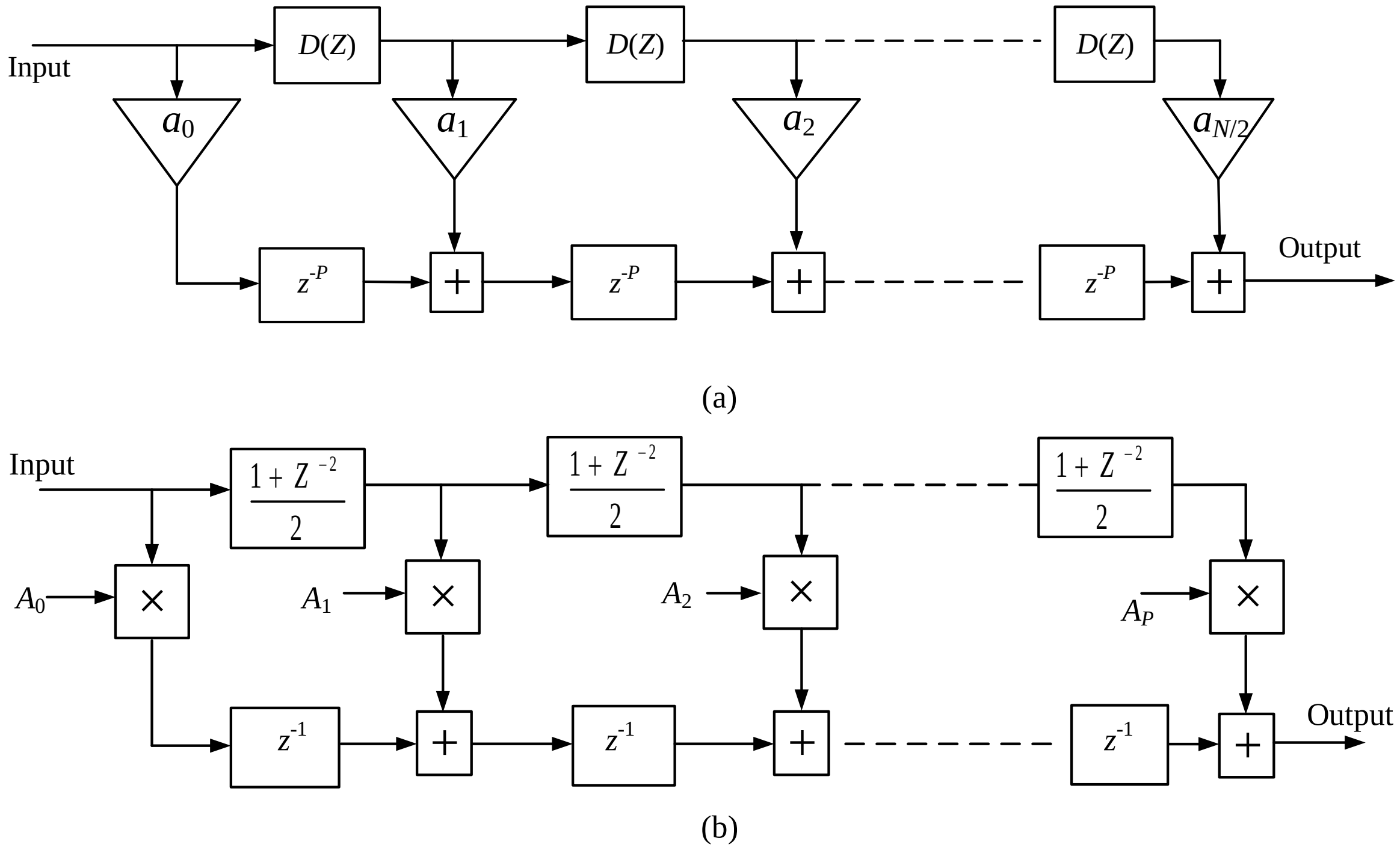

Fig. 14. (a) $P^{th}$-order frequency transformation based VCF filter implemented in Taylor structure. Coefficients $a_n$ and symmetric coefficients $h_n$ of the prototype filter of order $N$ are related via Chebyshev polynomials. (b) Structure of block $D(Z)$ for $P^{th}$-order transformation.

For the first-order frequency transformation [33], i.e., for $P = 1$, cutoff frequency of the frequency transformed filter can be either greater than or less than cutoff frequency of the prototype filter (with $\psi \geq 1$ or $\psi \leq 1$, respectively for the two cases). The cutoff frequency range is very limited and a lowpass prototype filter gives VCF lowpass filter. Prototype filter needs to be a bandpass filter to obtain VCF bandpass responses.

These drawbacks of the first-order transformation are overcome by second-order frequency transformation [34]. Second-order frequency transformation provides wider cutoff frequency range compared to that of the first-order frequency transformation and also provides smaller values for $\psi$. The cost incurred for this better performance is increase in the number of variable multipliers required and group delay. The maximum cutoff frequency range of the second-order frequency transformation based filter is obtained when $f_{c_proto} = 0.5$, and is approximately 27% of the Nyquist band and $\psi \leq 1$. The ratio $\psi$ increases as $f_{c_proto}$ deviates from 0.5. When $\psi > 1$, the prototype filter needs to be designed with narrower transition bandwidth, i.e. the prototype filter should be of higher order, so that the frequency transformed filter satisfies the desired final transition bandwidth specification. For $f_{c_proto} \neq 0.5$, if an additional constraint of $\psi \leq 1$ is imposed, then the cutoff frequency range of the frequency transformed filter decreases significantly. Lowpass prototype filter gives VCF lowpass filter and prototype filters needs to be a bandpass filter for VCF bandpass responses.

Wider cutoff frequency range is obtained in [35] by combining the second-order frequency transformations with the CDM-I technique. This filter needs a parallel add-delay chain for implementing the CDM-I operation. For frequency transformation and CDM-I based filter, $\psi > 1$ over a significant portion of its cutoff frequency range, and $\psi_{max}$ = 2.1. Hence, the prototype filter needs to be overdesigned (i.e. should have much higher order), resulting in significant increase in number of multipliers.

The modified second-order frequency transformation based filter (MSFT filter) [36] has two significant changes compared to the frequency transformation based filters. In MSFT filter, lowpass-to-highpass transformation is applied to the prototype filter before applying the second-order frequency transformation. Therefore, another lowpass response is obtained using complementary response of this frequency transformed highpass response. Also, the one-to-one mapping condition between the frequency variables is relaxed to obtain a two-band response. Two additional lowpass responses are obtained from this two-band response. Therefore, MSFT filter has much wider cutoff frequency range (from 0.0875 to 0.9125) compared to that of the second-order frequency transformation based filters, with the advantage that $\psi \leq 1$ [36].

The magnitudes of the coefficients $a_n$ of the Taylor structure increase exponentially with the increase in the filter order. Therefore, when the desired transition bandwidth is narrow, the fixed-point implementation of the frequency transformation based filters may not be feasible due to increase in the order of the filter. The interpolated second-order frequency transformation based filter (ISFT filter) [37] overcomes this limitation of the frequency transformation based filters. It provides even wider cutoff frequency range (approximately entire Nyquist band) along with narrower transition bandwidth. The prototype lowpass filter in the ISFT filter is designed using the second-order frequency transformations to provide continuously varying cutoff frequency on very small frequency range, with (relatively) wider transition bandwidth. Variable, overlapping ranges of cutoff frequency in the desired range, with narrower transition bandwidth, are obtained by making use of interpolation and masking technique.

Detailed architectures of these VCF filters are available in [35-37]. All these filters can provide variable lowpass, highpass, bandpass, and bandstop responses without any hardware re-implementation.

### 3.2.4 Spectral Parameter Approximation based VCF Filters

In the SPA technique [38-51], the frequency response of the variable (fractional delay or the cutoff frequency) filter is modelled as a polynomial function of the spectral parameter (fractional delay or the cutoff frequency), and the coefficients of this polynomial are the frequency responses of the fixed-coefficient sub-filters. In other words, the frequency response of the variable filter is expressed as a weighted sum of the frequency responses of the sub-filters and the weights are controlled by the spectral parameter (which, in the present context, is the cutoff frequency of the filter). Initially the SPA technique was developed to design the variable fractional delays, and was later adapted for the design of VCF filter.

The SPA based VCF FIR filter, implemented in the Farrow structure as shown in Fig. 15, consists of $L$+1 sub-filters, each of order $N$. Each sub-filter is a fixed-coefficient FIR filter same as in Fig. 2. The objective here is to find the optimal sub-filter coefficients such that the frequency response of the SPA based VCF filter approximates the desired VCF filter response as a function of $\alpha$. Various design methods utilizing the least-squares [38-43, 51] and/or minimax techniques [44-48],vector array decomposition [49], semi-infinite quadratic optimization [50] have been proposed in literature to solve the approximation problem by incorporating the desired peak to peak passband ripple and stopband attenuation constraints in the problem formulation. The optimal solutions can be obtained from the closed-form formulae [39-43, 47], or by solving the system of linear equations obtained by the discretization method [38, 44-46, 48-51]. In general, all the sub-filters in the Farrow structure have the same order, but unequal order case has also been treated in [44]. Traditionally, the approximation problem for the SPA based filter is formulated in the frequency domain [38-50]. In [51], a time-domain based approach is proposed which is shown to result in lower complexity SPA based filter when wide cutoff frequency range or narrow transition bandwidth is desired. For an SPA based filter, it is possible to obtain another lowpass response (and hence bandpass response by extension) using same sub-filters, just by adding extra ‘multiply-add’ chain corresponding to ‘$\alpha$’ multipliers to the filter shown in Fig. 15. For SPA based filters, $\psi = 1$.

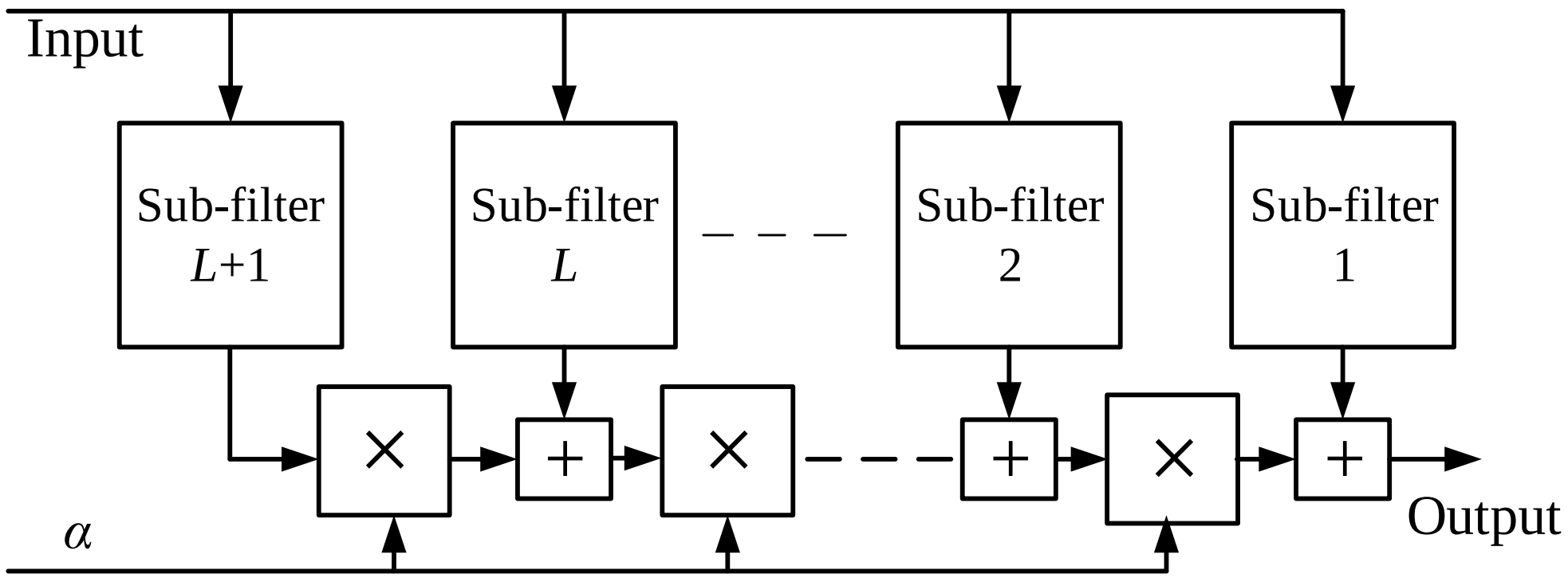

Fig. 15. Farrow structure implementation of SPA based filter. All sub-filters are fixed-coefficient FIR filters of equal order $N$.

However, all these SPA based filters have major limitations that their cutoff frequency range is very limited, and values of $N$, $L$, and the magnitudes of the sub-filter coefficients increase (resulting in large dynamic range) very rapidly as the desired cutoff frequency range increases and/or the desired transition bandwidth becomes narrow.

In [52], the FRM technique is modified by replacing the prototype and masking filters in the FRM technique by SPA based filters. The approximation problem is solved for each of these filters considering the frequency specifications given by the FRM design procedure. This combination of SPA and FRM technique results in extremely narrow transition bandwidth VCF filters. However, the complexity of this SPA-FRM technique is very high due to requirement of multiple SPA based filters, each of which is implemented using the Farrow structure. Furthermore, the passband ripples are very large and stopband attenuation is very poor in the SPA-FRM VCF filter and the other limitations of the SPA based filters, e.g. limited cutoff frequency range, large dynamic range of the sub-filter coefficients etc., also apply to this technique.

A solution to the limited range of the SPA based VCF filters was presented in [53], in which the SPA technique is combined with the MCDM-I technique to obtain continuous control over the cutoff frequency in the entire Nyquist band, with $\psi = 1$. The prototype SPA based filter in this SPA-MCDM VCF filter is designed for the limited cutoff frequency range of 0.25 to 0.5, and the filter responses in the remaining frequency range are obtained using the modified CDM-I technique. The SPA-MCDM VCF filter in [53] has lower complexity compared to the other SPA based filters on account of the requirement of limited cutoff frequency range. Still, when narrow transition bandwidth is desired along with small passband ripple and high stopband attenuation, the complexity of the SPA-MCDM VCF filter is very high.

An interpolated SPA (ISPA) based filter was proposed in [54] to extend the cutoff frequency range of the SPA based filters to entire Nyquist band, by integrating SPA and interpolation techniques. This method overcomes all the limitations of all the previously existing SPA based filters [38-53] and provides very wide cutoff frequency range and narrow transition bandwidth along with small passband ripple and high stopband attenuation. As the complexity of the SPA based filter increases with the increase in the desired cutoff frequency range, in the ISPA based filter, the SPA technique based prototype filter is designed to be variable only for very small cutoff frequency range, which is then extended over the desired (much wider) range by using the interpolation technique. Detailed architectures of these VCF filters are available in [52-54]. For all these filters, it is possible to obtain variable lowpass, highpass, bandpass, and bandstop responses without any hardware re-implementation.

A qualitative summary of the above mentioned techniques is provided in Table III. We note that in all the cases, lowpass VCF responses are obtained from lowpass prototype filter. In addition, bandpass responses can be obtained directly for all-pass transformation based filters, fractional delay structure + interpolation (IFDS) based filter, 2$^{nd}$ order frequency transformation + interpolation (ISFT) based filter, SPA + MCDM-I based filter, and SPA + interpolation (ISPA) based filter. As mentioned previously, a second multiply-add chain can be added to SPA based filter to obtain a second lowpass response, and a bandpass response can be obtained from two lowpass responses. For fractional delay structure based filter, a second add-delay chain can be implemented to obtain second lowpass response simultaneously. A frequency transformation based filter needs to be first implemented in transposed Taylor structure (as proposed in [68]) in order to use such a second add-delay chain. By extension, all these filters can be used to design a filter bank.

Table III: Qualitative comparison of VCF filters with *continuous* control over cutoff frequency.

| | Cutoff frequency range | *tbw* of VCF responses | Linear phase | Implementation complexity [a] | Number of variable multipliers [b] | Group delay [b] | Suitable for narrow *tbw* |
|---|---|---|---|---|---|---|---|
| 1$^{st}$ order all-pass transformation | Nyquist band | Variable | No | High | $N$ | *NA* | Yes |
| 1$^{st}$ order all-pass transformation + coefficient decimation | Nyquist band | Variable | No | Medium | $\geq N$ | *NA* | Yes |
| 2$^{nd}$ order all-pass transformation | Nyquist band | Variable | No | Very high | $2N$ | *NA* | Yes |
| Fractional delay structure based filter | Limited | Variable | Yes | Medium | $\geq N$ | $\geq N/2$ | Yes |
| Fractional delay structure + interpolation | Nyquist band, but disjoint ranges | Variable | Yes | Medium | $\geq N$ | $>> N/2$ | Yes |
| 1$^{st}$ order frequency transformation | Limited | Variable | Yes | Very high | $\geq N/2$ | $\geq N/2$ | No |
| 2$^{nd}$ order frequency transformation | Limited | Variable | Yes | Very high | $\geq N/2$ | $\geq N/2$ | No |
| 2nd order frequency transformation + interpolation | Nyquist band | Variable | Yes | Medium | ≥ $N/2$ | >> $N/2$ | Yes |
| SPA | Limited | Fixed | Yes | Very high | << $N$ | ≥ $N/2$ | No |
| SPA + MCDM-I | Nyquist band | Fixed | Yes | Very high | << $N$ | > $N/2$ | No |
| SPA + interpolation | Nyquist band | Variable | Yes | High | << $N$ | >> $N/2$ | Yes |

[a] Relative complexity for techniques considered in this table for same/similar $\{tbw, \delta_p, \delta_s\}_{_constraints}$. By complexity, we refer to total number of multipliers (variable and fixed) required to realize the filter.
[b] $N$ denotes the order of the variable-coefficient filter designed to obtain same cutoff frequencies with same $\{tbw, \delta_p, \delta_s\}_{_constraints}$. Group delay of this filter is $N/2$.
*NA*: not applicable.

In order to provide quantitative comparison, some of the linear-phase VCF filters from Table III are designed for $\{\delta_p, \delta_s\}_{_constraints}$ = {0.08 dB, -45 dB}. The corresponding results are presented in Table IV. These filters are able to provide a continuous control over the cutoff frequency along with narrow transition bandwidth. Even though we have used frequency-domain approach from [38] for designing prototype filter of the SPA + MCDM-I based filter, all the other frequency-domain approaches result in similar filter complexity. As can be noted, SPA + MCDM-I based filter has very high complexity when designed for narrower transition bandwidth specifications. ISFT and ISPA based filters provide trade-off between filter complexity and group delay.

Table IV: Quantitative comparison of VCF filters with *continuous* control over cutoff frequency.

| | Cutoff frequency range | *tbw* of VCF responses | Total number of multipliers | Maximum group delay (in samples) |
|---|---|---|---|---|
| 2nd order frequency transformation + interpolation based filter (ISFT filter) [37] | 0.0523 to 0.9477 | ≤ 0.035 | 188 | 600 |
| SPA + interpolation based filter (ISPA filter) [54] | 0.055 to 0.945 | ≤ 0.035 | 610 | 360 |
| SPA + MCDM-I based filter [53] with frequency-domain approach [38] based prototype filter | 0.025 to 0.975 | 0.05 | 2154 | 72 |
| SPA + MCDM-I based filter [53] with time-domain approach [51] based prototype filter | | | 1251 | 61 |

## 4. Conclusions

We have presented an all-inclusive review of VCF (variable cutoff frequency) FIR filter design techniques. The scope of this article was kept very specific and limited in order to remain coherent and make the reader comfortable in understanding the basics of VCF filters. Even though a large number of articles have been published till date, only a handful are mentioned for each design technique so as to introduce only the notable contributions. The other articles with similar or minor improvements over these can be easily retrieved from the references therein. A qualitative comparison of these techniques is provided in order to make the reader aware of their advantages and limitations. We have considered only the single-rate implementation. Multi-rate implementations of these filters, though they follow trivial procedures in most of the cases, could provide an additional research problem to interested parties.

An important aspect of digital filters is the implementation complexity. Deliberately, we have provided only the number of multipliers or multiplications as a measure of complexity for quantitative comparisons. We note that the actual hardware area requirement depends on the implementation strategy, platform and consideration for optimization algorithms during the design procedure. For instance, even though there is a significant difference in the number of multipliers required for ISFT and ISPA based filters, the actual hardware area could be comparable. This is because architecture of the SPA based filter provides sufficiently more opportunities for implementing MCM blocks and bit-widths for coefficients of frequency transformation based filters are significantly large compared to that of the SPA based filter. Similarly, due to the release of new hardware platforms having a powerful on-chip processors, it may be possible in some cases to generate the new coefficients for the variable-coefficient filter on-the-fly. Therefore, armed with the guidelines from this article, the reader is advised to make a detailed analysis and then choose the design technique on a case-by-case basis.